\documentclass[10pt]{article} 
\pdfoutput=1
\usepackage{amssymb,amsmath,amsfonts,theorem,subfigure} 
\usepackage{graphicx}
\usepackage{euler,palatino}
\usepackage{multirow}

\newcommand{\be}{\begin{equation}}
\newcommand{\ee}{\end{equation}}
\newcommand{\slz}{SL$(2,\mathbb Z)$}

\title{{\huge Critical exponents for the homology\\ of Fortuin-Kasteleyn clusters on a torus}}
\author{
{\Large Alexi Morin-Duchesne}\footnote{\ttfamily alexi.morin-duchesne{\char'100}umontreal.ca},\\
\it D\'epartement de physique\\ 
\it Universit\'e de Montr\'eal, C.P.\ 6128, succ.\ centre-ville, Montr\'eal\\ 
\it Qu\'ebec, Canada, H3C 3J7\\[10pt]
{\Large Yvan Saint-Aubin}\footnote{\ttfamily saint{\char'100}dms.umontreal.ca}\\
\it D\'epartement de math\'ematiques et de statistique\\ 
\it Universit\'e de Montr\'eal, C.P.\ 6128, succ.\ centre-ville, Montr\'eal\\ 
\it Qu\'ebec, Canada, H3C 3J7\\[10pt]}

\begin{document} 
\maketitle

%
%
 
\begin{abstract}
\noindent

A Fortuin-Kasteleyn cluster on a torus is said to be of type $\{a,b\}, a,b\in\mathbb Z$, if it possible to draw a curve belonging to the cluster that winds $a$ times around the first cycle of the torus as it winds 
$-b$ times around the second. Even though the $Q$-Potts models make sense only for $Q$ integers, they can be included into a family of models parametrized by $\beta=\sqrt{Q}$ for which the Fortuin-Kasteleyn clusters can be defined for any real $\beta\in (0,2]$. For this family, we study the probability $\pi({\{a,b\}})$ of a given type of clusters as a function of the torus modular parameter $\tau=\tau_r+i\tau_i$. We compute the asymptotic behavior of some of these probabilities as the torus becomes infinitely thin. For example, the behavior of $\pi(\{1,0\})$ is studied along the line $\tau_r=0$ and $\tau_i\rightarrow\infty$. Exponents describing these behaviors are defined and related to weights $h_{r,s}$ of the extended Kac table for $r,s$ integers, but also half-integers. Numerical simulations are also presented. Possible relationship with recent works and conformal loop ensembles is discussed.

\noindent Keywords: Fortuin-Kasteleyn clusters, torus, homology probabilities, homotopy probabilities, percolation, Ising model, logarithmic minimal models, SLE.
\end{abstract}

%
%

\newpage
\tableofcontents

%
%


\section{Introduction}
\label{sec:intro}

One of the main observables of two-dimensional percolation is the crossing probability between two disjoint subsets of the boundary of a domain. This domain is usually taken homeomorphic to a disk. As Langlands and his colleagues \cite{Langlands} were finishing their numerical study of universality and conformal invariance of crossing probabilities,
I.~Gelfand suggested to explore percolation on compact Riemann surfaces. The simplest surface to study is the torus and the most natural observable is then the homologic properties of the percolating cluster, or more precisely, the probability that a configuration contains a homologically non-trivial cluster. (Since these clusters are geometric objects, it might be easier to think about their homotopic properties instead 
of their homological ones.) Let $\omega_1$ and $\omega_2$ be the two-dimensional linearly independent vectors along the two  sides of the parallelogram defining the torus. In the following these will be identified to points in the complex plane. If a non-trivial cluster exists and if it winds $a$ times along $\omega_1$ of the torus while it wraps $b$ times along $-\omega_2$, the cluster is said to be of type $\{a,b\}$. All other non-trivial clusters of that configuration, if any, will be of the same type. (The integers $a$ and $b$ are coprimes. Types $\{a,b\}$ and $\{-a,-b\}$ are considered identical.) For that reason, the homology property of a configuration may be defined as the type of its non-trivial clusters. If the configuration contains no non-trivial cluster, it is said to be of type $\{0\}$. Finally, if the configuration contains a cluster that has both a path around the first cycle, that is along $\omega_1$, and a path along $\omega_2$, this configuration is of type $\mathbb Z\times \mathbb Z$. With that notation, each configuration is associated with one of the subgroups $H$ of the homology group $\mathbb Z\times\mathbb Z$ of the torus: $\{0\}, \mathbb Z\times \mathbb Z$ and $\{a,b\}$ with $a,b$ coprimes. The same notation $\{a,b\}$ is used for the type of a configuration and the subgroup generated by an element of that type. Langlands {\it et al} measured the probability of a few of these subgroups for percolation and gave some numerical evidence for their conformal invariance.

Pinson \cite{Pinson} obtained analytic expressions for the probability of these various subgroups as functions of the quotient $\tau$ of the fundamental periods $\omega_1,\omega_2\in\mathbb C$ of the torus. His computation relies on a clever argument giving an orientation to the curves bounding clusters. (See \cite{Coulomb,diFrancesco}.) This is done in a way that does not change the partition function, but does allow for the identification of the homology properties of intervening clusters. His computation is mathematically rigorous, except for the step taking the limit as the mesh goes to zero; for this, he used Nienhuis' renormalization group argument \cite{Coulomb} that ties the quantities under study to known results for the Coulomb gas. A more rigorous treatment of this step remains open.

Arguin \cite{Arguin} extended Pinson's argument to $Q$-Potts models, $Q=1,2, 3, 4$. To do so, he considered the Fortuin-Kasteleyn graphs or clusters of configurations. These are the natural extensions of the clusters of percolation, the Potts model with $Q=1$. Arguin showed that Pinson's formulae need only a small change for the $Q$-Potts model with $Q\ge2$. He also supported his new expression with numerical data for the four integer values of $Q$.

Works on or using probabilities of homology subgroups of FK clusters has not been limited to the theoretical predictions. Ziff, Lorenz, Kleban \cite{ZLK} were the first to provide numerical support for their universality. Later Newman and Ziff \cite{NZ} used them to give a precise estimate of the critical probability for site percolation on a square lattice. It was then the most precise available estimate. And recently they were again used to obtain precise estimates for critical probability for percolation on several lattices \cite{FDB}. (These probabilities are called {\em wrapping probabilities} in these works.)

In the definition of Potts models, $Q$ gives the number of states accessible to the basic random variables, often called spins. As such, $Q$ must be an integer. When the partition function is rewritten in terms of Fortuin-Kasteleyn graphs (hereafter FK graphs), the parameter $Q$ appears in the Boltzmann weight as $Q^{N_c}$ where $N_c$ is the number of FK connected components in the configuration. In this formulation, the condition that $Q$ be an integer may be relaxed. One then gets a one-parameter family of models, usually studied for the values of $Q$ in the interval $(0,4]$. It is between this family of models and the family of stochastic Loewner processes that a close tie seems to exist, and has been established for some particular cases. The stochastic Loewner equation with parameter $\kappa$ (SLE$_\kappa$) is believed to describe the growth of the boundary of a FK graph. The exact relationship between the two parameters $Q$ and $\kappa$ is
$$Q=4\cos^2 \frac{4\pi}\kappa$$
with $\kappa \in[4,8)$ and, again, $Q\in(0,4]$. Percolation corresponds to $\kappa =6$ (and $Q=1$) and the Ising model to $\kappa=\frac{16}3$ ($Q=2$). The mathematical tools to describe not only the boundary of a single FK cluster, but the set of loops described by the boundary of all clusters in a configuration are now emerging. {\em Conformal loop ensembles}, defined by Camia and Newman for percolation \cite{CamiaNewman} and more generally by Werner \cite{Werner} (see also \cite{doyon}), might allow for the rigorous study of homological properties of configurations, as defined and studied by Langlands {\it et al}, Pinson and Arguin. 

The goal of the present paper is to extract from the known expressions of the probabilities for the various homology subgroups their asymptotic behavior for two limiting cases. The first is when the quotient $\tau$ of the periods goes to infinity or to a real rational number. The second is when $Q$ goes to zero. 
The reason to study the latter is mostly curiosity. For the former, the reason is twofold. 
Many results proved using SLE techniques describe asymptotic behavior. The first reason is therefore to seek exponents to describe limiting behavior that might be easier to obtain with SLE (or CLE). The second reason is to probe deeper the relationship between SLE and conformal field theory (CFT). Several critical exponents appearing (rigorously) in the context of SLE had been predicted within CFT, and a large subset of these appeared in the Kac table of the associated minimal conformal model. It is agreed, but not proved, that SLE$_\kappa$ describes properties of the conformal theory with central charge
$$c(\kappa)=13-6\left(\frac{\kappa}4+\frac4\kappa\right).$$
Minimal models appear when $c$ and $\kappa$ are rational. Let $\kappa$ be rational and of the form $4p'/p$ with $p'>p\ge 1$, coprime integers. The conformal spectrum of the minimal model with central charge $c=c(\kappa)$ is constructed from the Virasoro highest weights 
\begin{equation}h_{r,s}=\frac{(\kappa r-4s)^2-(\kappa -4)^2}{16\kappa}, \qquad 1\le r\le p-1,\ \ 1\le s\le p'-1.
\label{eq:Virasoro}\end{equation}
It has been recognized however that the minimal models, constructed out of finite sets of primary fields and therefore of highest weights $h_{r,s}$, are probably too restrictive and might not capture all physical observables. Half-integers $r$ and $s$ have been considered~\cite{SaleurDuplantier} and several works about logarithmic minimal models have shown that the upper bounds on $r$ and $s$ need to be relaxed. (See, for example, \cite{mathieuRidout,PearceRasmussen} for recent arguments.) Maybe one of the most striking examples of this fact is Cardy's formula that describes the probability of crossing within a rectangle for percolation. For limiting geometries, that is for rectangles very wide or narrow, the probabilities approach $0$ or $1$ with the power of $h_{1,3}=\frac13$, an exponent that does not belong to the minimal set. Another example is related to the problem studied in the present note. In \cite{ArguinSaintAubin02}, Arguin and Saint-Aubin showed that, when the quotient $\tau$ of the fundamental periods of the torus tends to zero along the imaginary axis, the probability $\pi(\{1,0\})$ for the Ising model goes to $1$ as intuitively it should, but more precisely it goes as $\pi(\{1,0\})\rightarrow 1-(q^2)^{\frac18}f_1(q^2)-(q^2)^{\frac13}f_2(q^2)-\dots$ where $q=e^{i\pi \tau}$ and $f_1$ and $f_2$ are analytic in a neighborhood of $q=0$. The exponents are twice the highest weights $h_{1,2}=\frac1{16}$ and $h_{3,3}=\frac16$; the first belongs to the spectrum of the minimal model, the second does not. It is this observation that led us to ask whether exponents obtained by taking limits of the geometry would always be in the extended Kac table of the corresponding models when $\kappa$ is rational. (Every conformal weight $h_{r,s}$ is repeated an infinite number of times in the extended Kac table. Arguin and Saint-Aubin chose $(r,s)=(1,2)$ and $(3,3)$ for the leading exponents of the Ising model. We shall come back to this choice after determining the exponents for the general case.)

\begin{figure}
 \centering
  \includegraphics[width=0.60\textwidth]{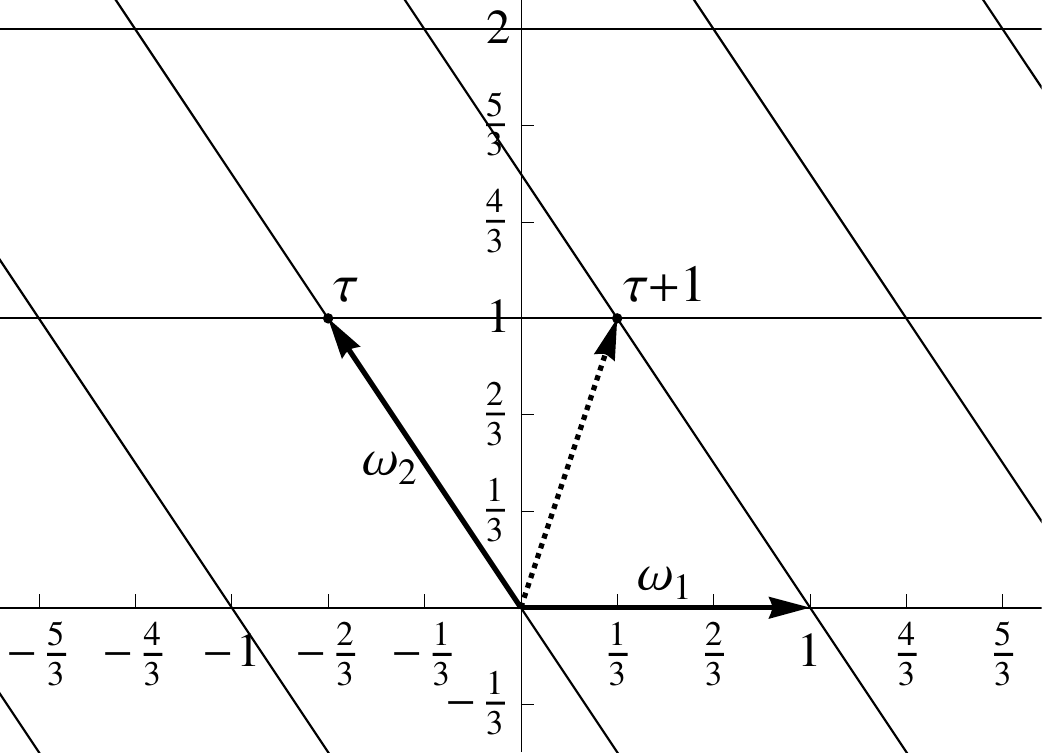}
 \caption{The torus in the complex plane, with $\tau = -2/3 + i$}
 \label{fig:complextorus}
\end{figure}
\begin{figure}[b!]
\begin{center}
\subfigure[]{\includegraphics[width=0.30\textwidth]{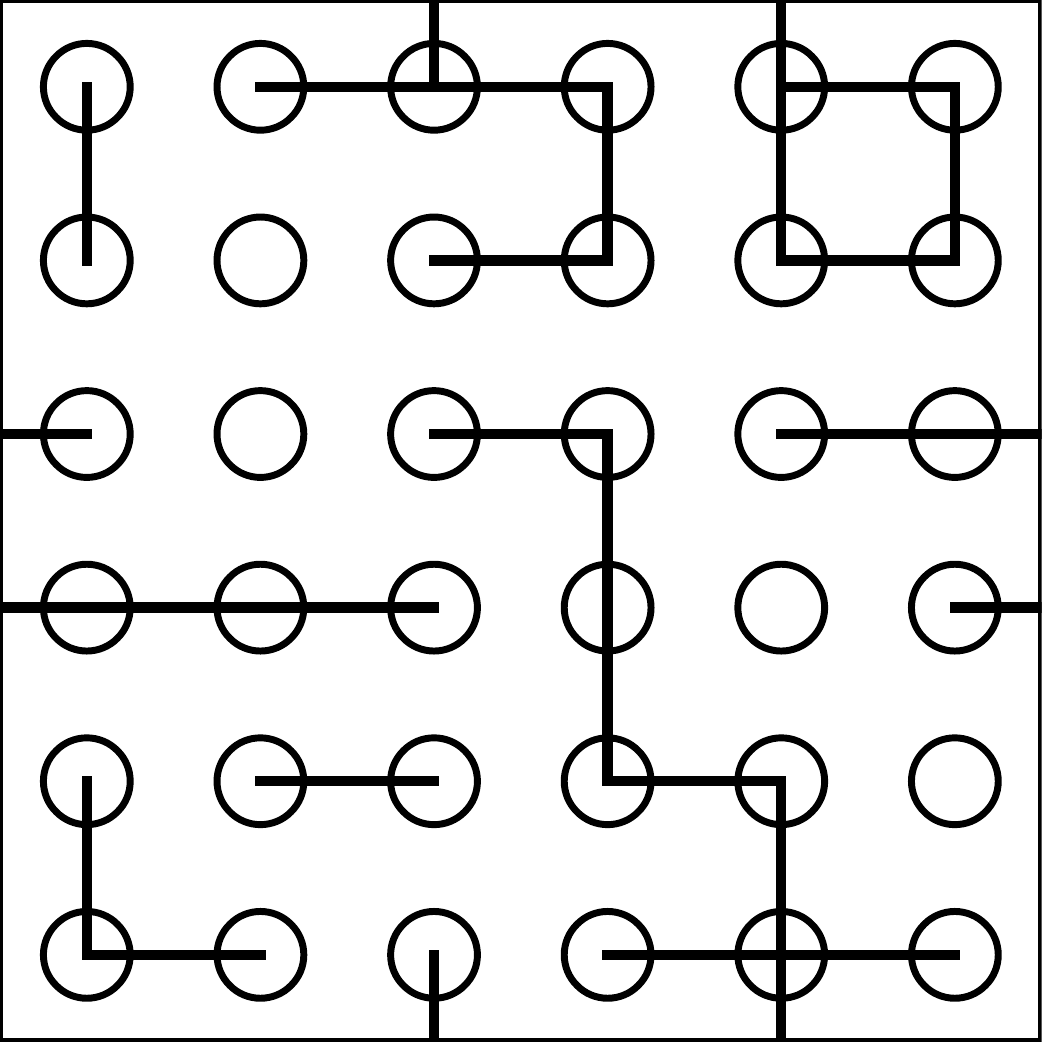}} \hfill
\subfigure[]{\includegraphics[width=0.30\textwidth]{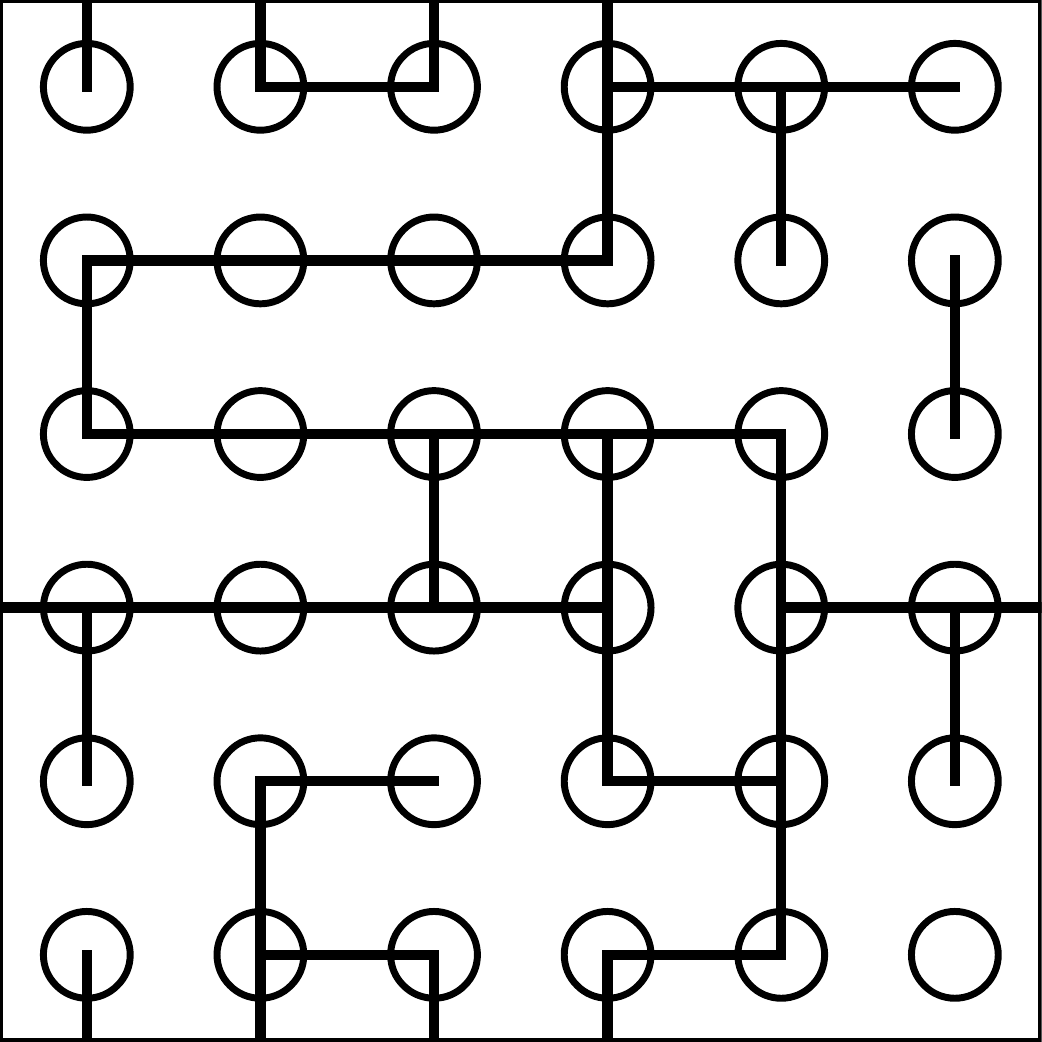}} \hfill
\subfigure[]{\includegraphics[width=0.30\textwidth]{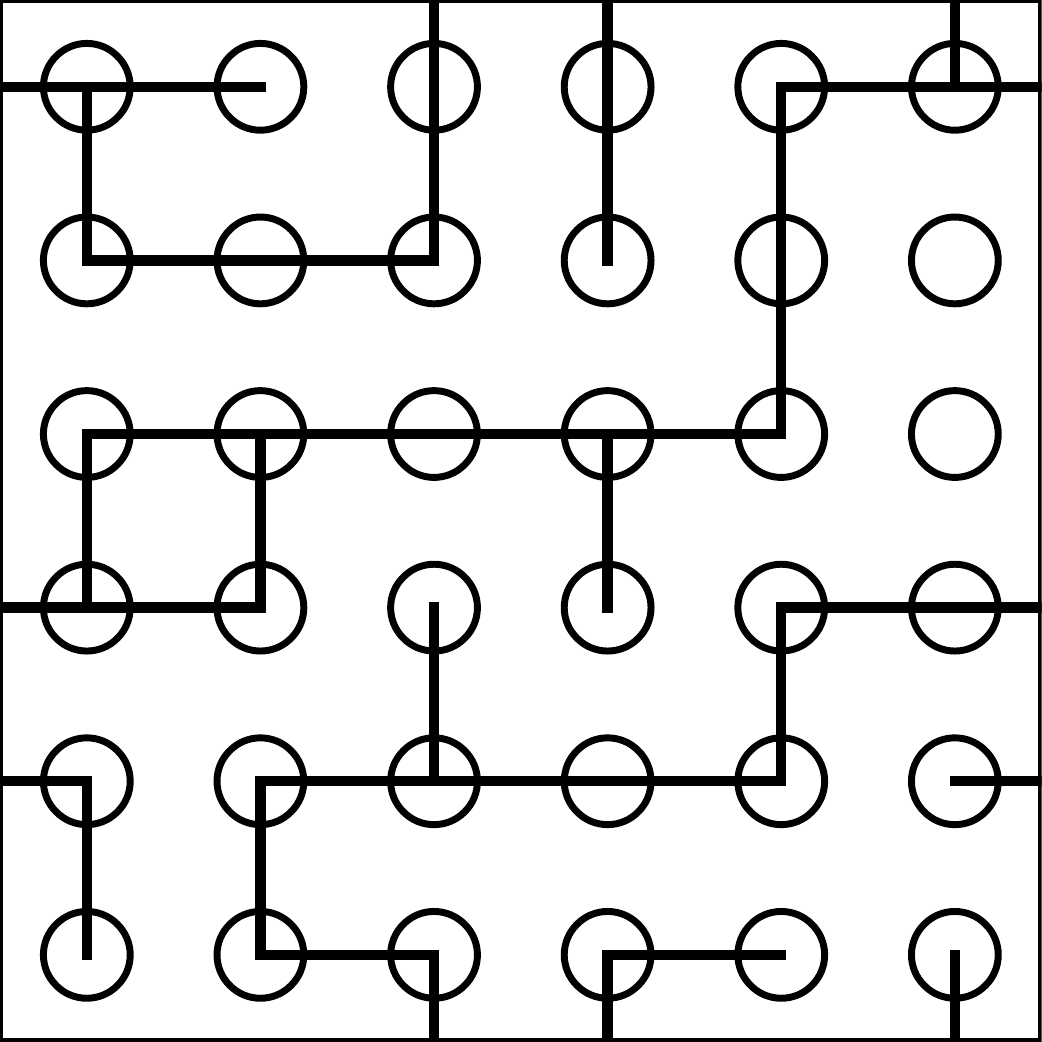}} \hfill
\caption{Examples of FK configurations of type (a) $\{0\}$, (b) $\mathbb{Z} \times \mathbb{Z}$ and (c) $\{2,-1\}$ groups, drawn on the torus with $\tau=i$}
 \label{fig:homotopygroups}
\end{center}
\end{figure}

Our notations are the following. The torus is identified with the quotient $\mathbb C/\{\omega_1,\omega_2\}$ where $\{\omega_1,\omega_2\}$ is the integral lattice generated by $\omega_1, \omega_2\in\mathbb C$ such that $0, \omega_1,\omega_2$ are not colinear. We choose $\omega_1=1$ and $\text{\rm Im\,} \omega_2>0$. Their quotient $\tau=\omega_2/\omega_1$ is the modulus of the torus with $\tau_r$ and $\tau_i>0$ its real and imaginary parts. Figure~\ref{fig:complextorus} shows these basic elements for a torus with $\tau=-\frac23+i$. We follow the convention set in \cite{Pinson, Arguin} for the winding numbers: they are positive in the direction of $\omega_1$ and $-\omega_2$. Figure~\ref{fig:homotopygroups} shows FK configurations of three different types drawn on the torus $\tau=i$. Configuration (c), for example, is of type $\{2,-1\}$ according to the above convention.

It is natural to break the partition function into sums 
over configurations of a given type or generating a given subgroup $H$. If $a\wedge b$ denotes the greatest common divisor of $a$ and $b$ (with $a\wedge0=a$ for all $a$), the partition function is
\begin{equation} \label{eq:Z}Z = Z(\{0\})+ Z(\mathbb{Z} \times \mathbb{Z}) +\sum_{a \wedge b=1}Z(\{a,b\}).\end{equation}
The observables under study are the probability of a given subgroup $H$, namely $\pi(H) = \frac{Z(H)}{Z}$. All these quantities depend on the size of the lattice covering the torus and the model labelled by $Q$. (For clarity we sometimes add an index, $Q$ or $\tau$, to quantities under study, e.g.~$Z=Z_Q$.) Their thermodynamic limit, when the mesh size goes to zero, are known at the critical temperature. The expressions obtained by Pinson~\cite{Pinson} for $Q=1$ and generalized by Arguin~\cite{Arguin} for $Q\in\{1,2,3,4\}$ are
\begin{align} Z(\{a,b\})&=\sum_{k\epsilon \mathbb{Z}}Z_{bk,ak}(g/4)(cos[\pi e_0k]-cos[\pi k]) \label{eq:Zab}\\
Z(\{0\})&={\textstyle{\frac12}}\sum_{m,m'\epsilon \mathbb{Z}}Z_{m,m'}(g/4)cos[\pi(m\wedge m')]  \label{eq:Z0}\\
Z(\mathbb{Z} \times \mathbb{Z})&=Q \times Z(\{0\})  \label{eq:ZZ2} 
\end{align}
where
\begin{equation} \label{eq:Zmm} Z_{m,m'}(g)= \frac{1}{|\eta(q)|^2}\sqrt{\frac{g}{\tau_i}}  e^{-\pi g|m\tau-m'|^2/\tau_i} \end{equation}
and
\begin{equation} Q = 4 cos^2[\pi e_0/2],\qquad g = 4-2e_0,\qquad e_0=2-8/\kappa, \qquad Q \in (0,4], \qquad e_0\in [0,1), \qquad \kappa\in[4,8). \end{equation}
The parameters $Q, g, e_0$ and $\kappa$ are all in one-to-one correspondence to one another in their respective range. (We use them in the way historical developments have introduced them.) Dedekind function is $\eta(q)=q^{1/24}\prod_{n\in\mathbb Z}(1-q^n)$.
Pinson's and Arguin's arguments extend trivially to the models of Fortuin-Kasteleyn cluster with a real $Q$ in the interval $(0,4]$. We use these expressions as our starting point.

The paper is organized as follows. In the next three sections, we study the following three limits: of $\pi(\{1,0\})$ when $\tau=i\tau_i$ and $\tau_i\rightarrow \infty$, of $\pi(\{a,b\})$ when $\tau=\frac cd+i\tau_i$ with $\tau_i\rightarrow 0$ and finally of $\pi(H)$ for any $H\subset \mathbb Z\times\mathbb Z$ when $Q\rightarrow 0$. The last section is devoted to Monte Carlo verifications of some of the results.

\section{The probability $\pi(\{1,0\})$ in the limit $\tau_r=0$,  $\tau_i \rightarrow \infty$}\label{sec:limit1}

The first limit to be studied is when $\tau = i\tau_i$ with $\tau_i \rightarrow \infty$, i.e.~the limit when the torus becomes a very thin ring. The corresponding parallelogram in the complex plane becomes an infinitely tall rectangle of constant width equal to $1$. Curves winding once along $\omega_1$ become very likely. In fact their relative length with respect to those winding once in the direction $\omega_2$ 
becomes negligible and it is therefore expected that, in this limit, all configurations will have curves of type $\{1,0\}$ and none of type $\{0,1\}$. In other words, $\pi(\{1,0\})\rightarrow 1$ and the probability of all other groups goes to $0$. What should be the expected behavior of $\pi(\{1,0\})$ for finite but very large $\tau_i$? Cardy's formula \cite{Cardy92} provides a fair guess. This formula gives, for percolation, the probability $\pi_h$ of horizontal crossing in a rectangle of width $H$ and height $V$ as a function of the aspect ratio $r=V/H$. For limiting geometries the probability behaves as
$$\pi_h(r)\underset{r\rightarrow0}{\longrightarrow}c_1e^{-\pi /3r}\qquad\text{\rm and}\qquad
1-\pi_h(r)\underset{r\rightarrow\infty}{\longrightarrow}c_2e^{-\pi r/3}$$
for known constants $c_1$ and $c_2$. Even though the intersections of a percolating cluster with the left and right edges of the rectangle might be in general at different height, these two intersections are likely to have points with the same vertical coordinates if the rectangle is very narrow, that is when $r\rightarrow 0$. Such a percolating cluster would be a FK cluster of type $\{1,0\}$, if opposite edges of the rectangle would be glued together. Therefore one may expect the following behavior
\begin{equation} \label{eq:crit} \pi(\{1,0\})= 1 - \sum_n c_nq^{\gamma_n} \end{equation}
with positive exponents $\gamma_n$ and the natural parameter $q = e^{-2\pi\tau_i}$ if the real part of $\tau$ vanishes. Note that $q$ goes to $0$ when $\tau_i\rightarrow \infty$. The goal of this section is to determine the leading exponents $\gamma_n$ as a function of $Q$ or, equivalently, $e_0$. (Some care should be exercised as the immediate extension of $q$ to a $\tau$ in the upper-half plane by $q=e^{2\pi i\tau}$ does not coincide with the usual definition of the nome of elliptic functions which is $e^{\pi i \tau}$.)

The probability $\pi(\{1,0\})$ is given in the form $Z_Q(\{1,0\})/Z_Q$. The first step is to express the numerator and denominator in a form suitable to extract these exponents. From \eqref{eq:Zab}:
$$Z_Q(\{1,0\})=\sum_{k\epsilon \mathbb{Z}}Z_{0k,1k}(g/4)(\cos[\pi e_0k]-\cos[\pi k]) =\frac{1}{|\eta(q)|^2}\sqrt{\frac{g}{4\tau_i}} \sum_{k\epsilon \mathbb{Z}}e^{-\frac{\pi g k^2}{4\tau_i}}(\cos[\pi e_0k]-\cos[\pi k]). $$
To rewrite the $e^{\frac{1}{\tau_i}}$ in terms of $q$, Poisson summation formula will be necessary:
\begin{equation} \label{eq:Poisson} \sum_{n\epsilon \mathbb{Z}}e^{-\pi an^2+bn} = \frac{1}{\sqrt{a}}\sum_{k\epsilon \mathbb{Z}}e^{-\frac{\pi}{a}(k+b/2\pi i)^2}.\end{equation}
After expanding the cosines in terms of exponentials, Poisson formula gives
\begin{equation}Z_Q(\{1,0\}) = \frac{1}{|\eta(q)|^2}\sum_{k\epsilon \mathbb{Z}} (q^{2(k+e_0/2)^2/g} - q^{2(k+1/2)^2/g}). \label{eq:zq10}\end{equation}
Since the function $q^{-1/24}\eta(q)$ has a Taylor expansion, the above form allows for the identification of the leading terms in the numerator. Note however that the expansion of $|\eta(q)|^2$ will not be used, since this same factor appears in the denominator. 

The denominator
$$Z_Q =(Q+1) Z(\{0\})+ \sum_{a\wedge b=1}Z(\{a,b\})$$
has two parts, which will be tackled separately. The partition function restricted to configurations with only trivial clusters is
$$Z_Q(\{0\})=\frac{1}{2|\eta(q)|^2}\sqrt{\frac{g}{4\tau_i}}\sum_{m,m'\in \mathbb{Z}}e^{-\frac{\pi g (m²\tau_i^2+m'²)}{4\tau_i}}\cos[\pi(m\wedge m')].$$
To get rid of the $cos[\pi(m\wedge m')]$, we notice that 
\begin{equation} \label{eq:resum} \sum_{m,m'\in \mathbb{Z}} = \sum_{m,m'\in 2\mathbb{Z}} +\  \Big(\sum_{m,m'\in \mathbb{Z}}-\sum_{m,m'\in 2\mathbb{Z}}\Big)\end{equation}
In the first sum, both $m$ and $m'$ are even which makes $m\wedge m'$ even and $\cos[\pi(m\wedge m')] = 1$. The other terms, in the parenthesis, are terms for which either $m$ or $m'$ is odd, and $\cos[\pi(m\wedge m')] = -1$. Therefore:
$$Z_Q(\{0\})=\frac{1}{4|\eta(q)|^2} \sqrt{\frac{g}{\tau_i}}\Big(2\sum_{m,m'\in 2\mathbb{Z}}-\sum_{m,m'\in \mathbb{Z}}\Big)e^{-\frac{\pi g (m\tau_i^2+m'^2)}{4\tau_i}}. $$
Sums over multiples of an integer $f\in \mathbb N$ will appear often and it is useful to define
\begin{align}\sigma(f,g)&=\sqrt{\frac{g}{\tau_i}}\sum_{m,m'\in f\mathbb{Z}}e^{-\frac{\pi g (m^2\tau_i^2+m'^2)}{4\tau_i}}\notag\\
&=\sqrt{\frac{g}{\tau_i}}(\sum_{m'\in \mathbb{Z}}e^{-\frac{\pi gf^2m'^2}{4\tau_i}})(\sum_{m\in \mathbb{Z}}e^{-\frac{\pi gf^2m^2\tau_i}{4}})\notag\\
&=\frac{2}{f}\sum_{m,m'\in \mathbb{Z}}q^{\frac{2m'^2}{gf^2} + \frac{gf^2m^2}{8}}
\label{eq:sigma}
\end{align}
where Poisson formula \eqref{eq:Poisson} was used again in the last line. The partition function $Z_Q(\{0\})$ is then
\begin{equation} \label{eq:zed0}Z_Q(\{0\}) = \frac{1}{4|\eta(q)|^2} (2\sigma(2,g)-\sigma(1,g)) = \frac{1}{2|\eta(q)|^2} \sum_{m,m'\in \mathbb{Z}}(q^{\frac{m'^2}{2g} + \frac{gm^2}{2}} - q^{\frac{2m'^2}{g} + \frac{gm^2}{8}}). \end{equation}

The remaining term of $Z_Q$, that includes configurations with non-trivial FK clusters of type $\{a,b\}$ for all $a$ and $b$ coprimes, is more complicated. The sum 
\begin{equation} \label{eq:sumab} \sum_{a\wedge b=1}Z_Q(\{a,b\}) = \sum_{m,m'\in \mathbb{Z}}Z_{m,m'}(g/4)(\cos[\pi e_0 (m\wedge m')]-\cos[\pi (m\wedge m')]) 
\end{equation}
contains two terms. The second with $\cos[\pi (m\wedge m')]$ is exactly twice the partition function $Z_Q(\{0\})$ just calculated. The first with $\cos[\pi e_0 (m\wedge m')]$ does not simplify as easily; the sums must be reorganized before~\eqref{eq:Poisson} is used. To do so, consider, for $m$ fixed, the function $\cos[\pi e_0(m\wedge m')]$. When $m$ is non-zero, it is periodic in $m'$ with period $m$. Therefore 
\begin{equation} \sum_{m' \in \mathbb{Z}}Z_{m,m'}(g/4) \cos[\pi e_0(m\wedge m')] = 
\sum_{d|m}\sum_{m'\in d\mathbb{Z}} C(d,e_0) Z_{m,m'}(g/4), \qquad m\neq 0 \label{eq:fmmp}
\end{equation} 
with 
\begin{equation} \label{eq:C}
C(d,e_0) =\sum_{d_2|d}\cos(d_2 \pi e_0)\mu(\frac{d}{d_2}) \end{equation}
where $\mu(x)$ is the M\"{o}bius function of $x$. (Recall that $\mu(1)=1$, $\mu(n)=0$ if $n$ has repeated prime factors and $\mu(n)=(-1)^\ell$ if $n$ is the product of $\ell$ distinct primes.) To get (\ref{eq:fmmp}-\ref{eq:C}), the sum over  $m'$ was divided into sums over subsets which have the same value of $\cos[\pi e_0(m\wedge m')]$, in a fashion similar to the splitting proposed in equation~\eqref{eq:resum}. These subsets are closely related to the divisors of $m$, therefore leading to the splitting into sums over the multiples of these divisors. We must stress, however, that the only divisors to be considered in ${d|m}$ are the positive ones.
The remaining sum can be written with the help of \eqref{eq:fmmp} as
\begin{align*}\sum_{m,m'\in \mathbb{Z}}Z_{m,m'}(g/4)\cos[\pi e_0 (m\wedge m')]&
=\sum_{m'\in \mathbb{Z}} Z_{0,m'}(g/4)\cos(\pi e_0m') 
 +\sum_{m\in \mathbb{Z^*}} \sum_{d|m}\sum_{m'\in d\mathbb{Z}} C(d,e_0) Z_{m,m'}(g/4)\\
&=\sum_{m'\in \mathbb{Z}} Z_{0,m'}(g/4)\cos(\pi e_0m') 
 + \sum_{m\in \mathbb{Z^*}} \sum_{d|m}\sum_{m'\in \mathbb{Z}} C(d,e_0) Z_{m,dm'}(g/4)
\end{align*}
where $\mathbb Z^*=\mathbb Z\setminus \{0\}$.
In the above expression, the terms with $m=0$ get a special treatment because of the particular definition of $m\wedge m'$ when $m$ is $0$. 
These were already encountered in the computation of $Z_Q(\{1,0\})$ and are equal to
$$\sum_{m'\in \mathbb{Z}} Z_{0,m'}(g/4)\cos(\pi e_0m')= \frac{1}{|\eta(q)|^2}\sum_{k\epsilon \mathbb{Z}} q^{2(k+e_0/2)^2/g}.$$
For the triple sum, the sum over divisors can be rearranged using
 $$\sum_{m\in \mathbb{Z^*}}\sum_{d|m} h(m,d) = \sum_{d\in \mathbb{N^*}}\sum_{m\in d\mathbb{Z^*}}  h(m,d) = \sum_{m\in \mathbb{Z^*}} \sum_{d\in \mathbb{N^*}} h(md,d)$$
and similarly for the sum of ${d_2|d}$ in $C(d,e_0)$. These manipulations have doubled the number of sums in~\eqref{eq:sumab} from two, on $m$ and $m'$, to four, on $m, m', d, d_2$. This is the price to pay to use Poisson formula on the sum over $m'$ and cast everything into powers of $q$. The result is
\begin{align}  \sum_{m\in\mathbb Z^*,m'\in \mathbb{Z}}Z_{m,m'}(g/4)&\cos[\pi e_0 (m\wedge m')]\notag \\
&=\sum_{m\in \mathbb Z^*}\sum_{d\in\mathbb N^*}\sum_{m'\in\mathbb Z} C(d,e_0) Z_{dm,dm'}(g/4)\label{eq:term1v2}\\
&=  \frac{1}{|\eta(q)|^2} \sum_{m \in \mathbb{Z^*}}\sum_{d,d_2\in \mathbb{N^*}}\sum_{m'\in \mathbb{Z}} \frac{\cos(\pi e_0d_2) \mu(d)}{d d_2} q^{\frac{g(md d_2)^2}{8}+\frac{2m'^2}{g(d d_2)^2} }\label{eq:term1} \end{align}
and the complete partition function $Z_Q$ is
\begin{align}|\eta(q)|^2 Z_Q = \displaystyle\sum_{k\in \mathbb{Z}}q^{2(k+e_0/2)^2/g}& + \frac{(Q-1)}{2}\displaystyle\sum_{m,m'\in \mathbb{Z}}(q^{\frac{m'^2}{2g} + \frac{gm^2}{2}} - q^{\frac{2m'^2}{g} + \frac{gm^2}{8}})\notag\\
&+\displaystyle\sum_{\substack{m\in \mathbb{Z^*}\\d,d_2\in \mathbb{N^*}\\m'\in \mathbb{Z}}}\frac{\cos(\pi e_0d_2) \mu(d)}{d d_2} q^{\frac{g(md d_2)^2}{8}+\frac{2m'^2}{g(d d_2)^2}}.\label{eq:leZQ}
\end{align}
The probability $\pi(\{1,0\})$ is the quotient of $Z_Q(\{1,0\})$ given in~\eqref{eq:zq10} and of $Z_Q$.

It is now straightforward to see that the lowest-order term in $q$ is $\frac{e_0^2}{2g}$ for both the denominator $Z_Q$ and the numerator $Z_Q(\{1,0\})$. After simplification of the common factor $q^{{e_0^2}/{2g}}$, an expansion can be done to obtain the whole sets of exponents. An exhaustive list of possible exponents is given by taking exponents in the numerator and in the denominator, plus any integral linear combinations of them which arise from higher order terms in the expansion. The possibility that some of them could have vanishing coefficients is not excluded.

It is interesting to compare the leading exponents with values \eqref{eq:Virasoro} given by CFT in the Kac table~\cite{FMS}. In terms of $g$ and $e_0$ they are  
\begin{equation} \label{eq:kac}h_{r,s}=\frac{[r-(g/4)s]^2-e_0^2/4}{g} \end{equation}
for $r,s$ positive integers. Note that $\frac{e_0^2}{4g}$ is half the power of $q$ that was substracted to simplify the numerator and denominator. The first exponents for $\pi(\{1,0\})$ are given by
\begin{equation} \label{eq:exponent1}\gamma_1 = \frac{1-e_0^2}{4(2-e_0)}, 
\qquad\gamma_2 = \frac{1-e_0}{2-e_0} \end{equation}
and their integer multiples.
On the range of $e_0$, $\gamma_2>\gamma_1$. The two exponents become equal in the limit $e_0=1$ ($Q=0$); this particular case will be studied in section~\ref{sec:Q0}.

Coincidences of these leading exponents or higher ones with elements from the Kac table, 
if any, will occur in the form $\gamma=2h_{r,s}$ for some $r,s$ because of the contribution of holomorphic and anti-holomorphic sectors. Such coincidences do occur. The simplest $r$ and $s$ giving $\gamma_1$ are $r=\frac12,s=0$ and, those giving $\gamma_2$, $r=0,s=1$. It is somewhat unusual to choose vanishing $s$ or $r$. Recall however that, for logarithmic minimal models, the Kac table is extended and the periodicity of elements $h_{r,s} = h_{r+p,s+p'}$ for the model with $\kappa=4p'/p$ allows to choose $r$ and $s$ positive. For some minimal models, it is however impossible to account for $\gamma_1$ with {\em integers} $r$ and $s$. Half-integers must be used. Arguin and Saint-Aubin \cite{ArguinSaintAubin02} identified the two leading exponents for the Ising model to $2h_{1,2}$ and $2h_{3,3}$. Note that, when either $r$ or $s$ is zero, then $h_{r,s}=h_{-r,-s}$. Moreover, if half-integer indices are included, the periodicity property can be refined to $h_{r,s}=h_{r+p/2,s+p'/2}$. The Ising model corresponds to $p=3, p'=4$ and their exponents are related to ours by $h_{1,2}=h_{-1/2,0}=h_{1/2,0}$ and $h_{3,3}=h_{0,-1}=h_{0,1}$.

These two exponents $\gamma_1$ and $\gamma_2$ are related to the fractal dimensions of geometric objects, namely the mass and the hull of a cluster respectively. (See \cite{Stanley, Duplantier, SaleurDuplantier}. For an extension of these geometric objects to loop gas models, see \cite{ysappjr}.) In the FK formulation of the $Q$-Potts models, the FK cluster mass attached to a site is the number of bonds in the component of the FK graph containing this site. In the plane, the hull of a FK cluster is the set of bonds that can be reached from infinity without crossing any bond from the cluster. (On a torus, each cluster has an inner and an outer hull.) Their fractal dimension is $2-2\Delta$ where $\Delta$ is $h_{1/2,0}$ for the cluster mass and $h_{0,1}$ for the hull. 

A natural explanation for $h_{1/2,0}$ in the present context is provided by Cardy \cite{Cardy98} (see also \cite{SaleurDuplantier}). Note first that the only way to keep a configuration from having a cluster of type $\{1,0\}$ is to have a cluster in the vertical direction. It is likely that its type will be $\{m,1\}$ for some $m\in\mathbb Z$ or $\mathbb Z\times \mathbb Z$. Cardy gives an expression for the probability $P(n,k)$ of having $n$ clusters connecting the two extremities of a cylinder whose length is $k$ times the perimeter of its section. He finds $\log P(n,k)\sim -\frac{2\pi}3(n^2-\frac14)k$ if $n\ge 2$. He points out that this expression evaluated at $n=1$ is {\em not} the probability of having a single cluster between the two extremities, but it is the probability of having a single cluster between the extremities that does not wind in the other direction.
When all configurations with a single cluster are considered, disregarding their behavior in the other direction, the probability is larger and given by $\log P(1,k)\sim -\frac{5\pi}{24}k$. Because $e_0=\frac23$ for percolation, our first correction term is $q^{\gamma_1}=e^{-5\pi\tau_i/24}$, in agreement with his result.

In a recent study of percolation, Ridout \cite{Ridout} has argued that the primary field responsible for changing boundary conditions in the computation of Watts' formula should be $\phi_{2,5/2}$. This identification forces him to shift, in the extended Kac table, the admissible values of $s$ by $\frac12$ when $r$ is even. One would like to see a relationship with our identification of $\gamma_1$ as $2h_{1/2,0}$. However it is $r$ that takes an half-integer value in our case, and $s$ in his case. Moreover the value $h_{1/2,0}=\frac{5}{96}$ does not appear in his shifted extended Kac table.

The other exponents in the numerator of $\pi(\{1,0\})$ are also part of the extended Kac table. They appear with $r=k+1,s=1$ for $\frac{2(k+e_0/2)^2-e_0^2/4}{g}$ and $r = k+1/2,s=0$ for $\frac{2(k+1/2)^2-e_0^2/4}{g}$. Not all the exponents of the denominator however appear in the extended Kac table, even if one allows half-integers $r$ or $s$. For example the denominator of the exponents $\frac{2m'^2}{g(d d_2)^2}$ appearing in the last sum of $Z_Q$ is not bounded. There is no hope to find them all in the extended Kac table. Could these terms drop out of the sum because of cancellations? The general case is difficult to assess, but this happens in simple cases.

It is indeed possible to find simpler form for the denominator for the four integral values $Q=1,2, 3, 4$ that correspond to $e_0=\frac23,\frac12,\frac13, 0$. For $d>0$ and these particular values of $e_0$ and for $e_0=1$, the function $C(d,e_0)$ is particularly simple:
\begin{align}
C(d,0)&= \delta_{d,1}\label{eq:c0}\\
C(d,\frac{1}{3})&= \frac{\delta_{d,1}}{2} -\delta_{d,2} - \frac{3\delta_{d,3}}{2} + 3\delta_{d,6}\\
C(d,\frac{1}{2})&= 2\delta_{d,4} - \delta_{d,2} \\
C(d,\frac{2}{3})&= \frac{3\delta_{d,3}}{2} -  \frac{\delta_{d,1}}{2}\\
C(d,1)&= 2\delta_{d,2} - \delta_{d,1}.\label{eq:c1}
\end{align}
The proof of these formulae is given in the Appendix. The contribution of
$\sum_{m, m'\in\mathbb Z}Z_{m,m'}(g/4)\cos[\pi e_0 (m\wedge m')]$ to $Z_Q$ is then much simpler. It is 
\begin{equation*}
\frac12\sum \left(q^{\frac{m'^2}{12}+3m^2} - q^{\frac{3m'^2}{4}+ \frac{m^2}{3}}\right)\quad \text{\rm for\ }Q=1, \qquad
\frac12\sum \left(q^{\frac{m'^2}{24}+6m^2} - q^{\frac{m'^2}{6}+ \frac{3m^2}{2}}\right)\quad \text{\rm for\ }Q=2,
\end{equation*}
\begin{equation*}
\frac12\sum \left(q^{\frac{m'^2}{60}+15m^2} - q^{\frac{m'^2}{15}+ \frac{15m^2}{4}} -  q^{\frac{3m'^2}{20}+\frac{5m^2}{3}} + q^{\frac{3m'^2}{5}+ \frac{5m^2}{12}}\right)\quad \text{\rm for\ }Q=3
\end{equation*}
\begin{equation*}
\text{\rm and}\qquad\sum q^{\frac{m'^2+m^2}{2}}\quad \text{\rm for\ }Q=4.
\end{equation*}
All the sums above are on $m, m'\in\mathbb Z$. It is then straigthforward to show that, upon simplification of the factor $q^{e_0^2/2g}$, these forms (and therefore $Z_Q$) can be written as a product $\sum_{i,j}(q^{2h_i}f_i(q^2))(q^{2h_j}f_j(q^2))$ of two {\em finite} sums where $f_i,f_j$ are analytic in a neighborhood of $0$ and where all the $h_i,h_j$ belong to the corresponding extended Kac table for some $r$ and $s$ integers
in the range $[0,p]$.

\section{Probabilities $\pi(\{a,b\})$ in the limit $\tau _r=c/d$,  $\tau _i \rightarrow 0$}
\label{sec:limit2}

\begin{figure}
 \centering
  \includegraphics[width=0.60\textwidth]{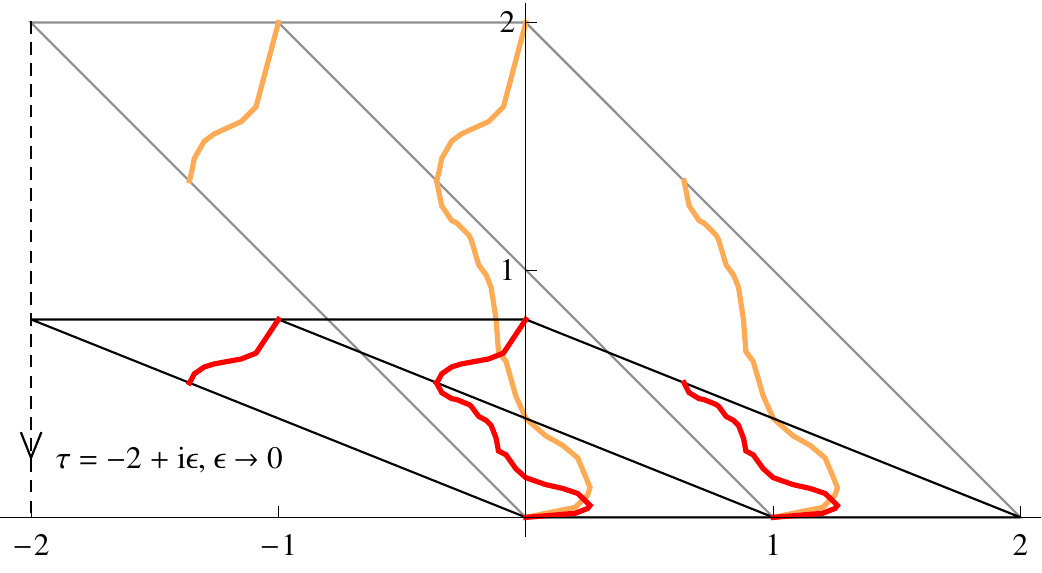}
 \caption{Curves of type $\{-2,1\}$ become more likely as $\tau$ approaches $\tau \sim -2 + i0^+$}
 \label{fig:limit2}
\end{figure}

The probabilities $\pi(\{a,b\})=\pi_\tau(\{a,b\})$ also have a limit, either $0$ or $1$, when $\tau$ approaches a rational number on the real line. To see this first intuitively, consider the case $\{a,b\}=\{-2,1\}$. Figure \ref{fig:limit2} presents two tori whose modulus parameter is on the line $-2+i\tau_i$. For each, two neighboring fundamental parallelograms have been drawn. A curve linking the origin to the vertex at $z=2i\tau_i$, like those shown, is of type $\{-2,1\}$. A curve of type $\{-2,1\}$ does not need to start at a vertex, of course, but those drawn show how the curves of this type will be prevailing. Indeed, as the modulus parameter $\tau$ slides down the vertical line $\tau_r=-2$, these curves become very short and likely. Therefore the probability $\pi_\tau(\{-2,1\})$ should converge to a number larger than $0$ when $\tau\rightarrow -2$. This section shows that it actually goes to $1$.

We have identified a torus with its modulus $\tau$, a complex number in the upper-half plane $\mathbb H$. As it is well-known, this correspondence is not unique, since any pair $\omega_1'=m\omega_1+n\omega_2$ and $\omega_2'=p\omega_1+q\omega_2$ with $m,n,p,q\in\mathbb Z$ and $mq-np=1$ describes the same torus, but with a new modulus $\tau'=\omega_2'/\omega_1'$. The special linear transformations $\left(\begin{smallmatrix}q & p\\ n& m\end{smallmatrix}\right)$ with integer coefficients
and determinant $1$
form the modular group \slz. It is generated by two matrices
$$s=\begin{pmatrix} 0 & -1\\ 1 & 0\end{pmatrix}\qquad\text{\rm and}\qquad 
t=\begin{pmatrix} 1&1\\ 0&1\end{pmatrix}$$
whose action on $\tau$ is 
\begin{equation}\label{eq:actionslz}
\tau\overset{s}{\mapsto} -1/\tau\qquad\text{\rm and}\qquad
\tau\overset{t}{\mapsto}\tau +1.
\end{equation}
The probabilities $\pi(\{0\})=\pi_\tau(\{0\})$ and $\pi(\mathbb Z\times\mathbb Z)=\pi_\tau(\mathbb Z\times\mathbb Z)$ are invariant under the change of $\tau$ by an element
of \slz, but the probabilities $\pi_\tau(\{a,b\})$ are not. Arguin \cite{Arguin} gave their transformation laws
\begin{equation} \label{eq:piabtransf}\pi_{\tau}(\{a,b\}) = \pi_{\tau+1}(\{a+b,b\}) =  \pi_{-1/\tau}(\{-b,a\}) \end{equation}
or, equivalently
\begin{equation}\label{eq:piabtransf2}\pi_{\tau}(\{a,b\})=\pi_{g\tau}(g\cdot\{a,b\}),
\qquad g\in \text{\rm SL}(2,\mathbb Z)
\end{equation}
 
where $\tau\mapsto g\tau$
denotes the action defined by \eqref{eq:actionslz} and $g\cdot\{a,b\}$ stands for the matrix multiplication $g\left(\begin{smallmatrix}a\\ b\end{smallmatrix}\right)$. These transformations follow immediately from the form \eqref{eq:Zab} of the partition function $Z_Q(\{a,b\})$.

A simple application of the modular transformation gives $\pi(\{0,1\})$ in terms of $\pi(\{1,0\})$, namely $\pi_\tau(\{0,1\})=\pi_{-1/\tau}(\{1,0\})$. The result of the previous section implies easily that
$$\pi_{\tau=0+i\tau_i}(\{0,1\})=\pi_{\tau=0+i/\tau_i}(\{1,0\})=\left.\frac{Z_Q(\{1,0\})}{Z_Q}\right|_{q'}$$
where the partition functions are evaluated at $q'=e^{-2\pi/\tau_i}$. The limiting behavior $\tau_i\rightarrow 0^+$ will therefore be characterized by the same exponents obtained for $\pi(\{1,0\})$ when $\tau_i\rightarrow \infty$.

Let $g\in\text{\rm SL}(2,\mathbb Z)$ and $z\mapsto gz$ the associated map. It is conformal, one-to-one on $\mathbb H$ and maps the real line onto itself. The image under such a map of the imaginary axis will therefore be a circle intersecting the real axis at right angles. 

Let $\{a,b\}$ be a pair of coprime integers. Then there are integers $p$ and $q$ such that $pa+qb=1$. Therefore $g=\left(\begin{smallmatrix} a&-q\\ b&p\end{smallmatrix}\right)\in\text{\rm SL}(2,\mathbb Z)$. The action of $g$ on $\mathbb H$ maps a point $\tau=i\tau_i, \tau_i>0$, on the positive imaginary axis into the point
\begin{equation}\label{eq:imageSousG}
i\tau_i\mapsto \frac ab\left(\frac{1-pq/ab\tau_i^2}{1+p^2/b^2\tau_i^2}\right)+\frac{i}{\tau_ib^2}\left(\frac1{1+p^2/b^2\tau_i^2}\right).
\end{equation}
The two parentheses behaves as $1+O(\tau_i^{-2})$ for $\tau_i\rightarrow \infty$. This repeats the statement just made: the image of the positive imaginary axis intersects the real line at right angles. The two intersection points are the image of $\tau=0$ and $\infty$. Note that, even though the solution $p, q$ of $pa+qb=1$ is not unique, the form of $g$ was chosen so that the image of $\tau=\infty$ and the tangent at this point do not depend of the pair $p,q$, but only on $a,b$.

For this particular element $g\in\text{\rm SL}(2,\mathbb Z)$, the modular transformation of the probability $\pi_\tau(\{1,0\})$ gives
$$\pi_\tau(\{1,0\})=\pi_{(a\tau-q)/(b\tau+p)}(\{a,b\}).$$
Because of \eqref{eq:imageSousG} the behavior of $\pi_\tau(\{1,0\})$ for $\tau=i\tau_i$ with $\tau_i\rightarrow \infty$ fixes the behavior of $\pi_{\tau'}(\{a,b\})$ at $\tau'=\frac ab+\frac{i}{\tau_ib^2}$. More precisely
\begin{equation}\label{eq:oscillo}
\pi_{\frac ab+i\epsilon}(\{a,b\})=1-\sum_n c_nq^{\gamma_n/b^2}, \qquad \text{\rm with\ }
q=e^{-2\pi/\epsilon}
\end{equation}
with the same $c_n$ and $\gamma_n$ as in \eqref{eq:crit}. Consequently all others $\pi(\{c,d\})$ with $\{c,d\}\neq \pm \{a,b\}$ should go to zero when $\tau\rightarrow \frac ab$.

\begin{figure}
 \centering
  \includegraphics[width=0.8\textwidth]{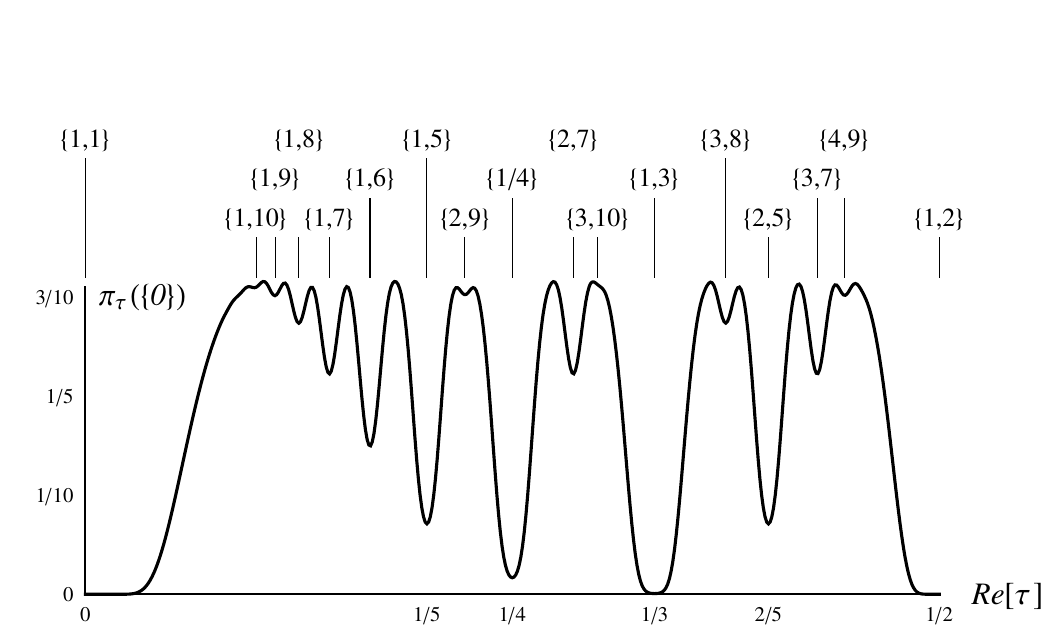}
 \caption{The function $\pi_\tau(\{0\})$ as a function of $\tau_r$ for $\tau_i=\frac1{100}$. Each valley of the graph is labeled by the subgroup $\{a,b\}\subset \mathbb Z\times\mathbb Z$ whose probability tends to $1$ at $\tau_r=a/b$.}
 \label{fig:oscillo}
\end{figure}

Ziff, Lorenz and Kleban \cite{ZLK} noticed that the probability $\pi_\tau(\mathbb Z\times \mathbb Z)$, and therefore $\pi_\tau(\{0\})$, develop oscillations as a function $\tau_r$ when $\tau_i$ is close to zero. Their qualitative observation is made quantitative by \eqref{eq:oscillo}. Figure \ref{fig:oscillo} draws the function $\pi_\tau(\{0\})$ as a function of $\tau_r\in[0,\frac12]$ for $\tau_i=\frac1{100}$. (It is sufficient to restrict the domain of $\tau_r$ to $[0,\frac12]$ as $\pi_\tau(\{0\})=\pi_{\tau+1}(\{0\})$ and the function $f(\tau_r)=\pi_{\tau+\frac12}(\{0\})$ is even for a fixed value of $\tau_i$.) The oscillatory behavior is obvious. Each valley of the graph occurs when $\tau_r$ is a simple fraction $a/b$, for coprime integers $a$ and $b$ and $b\le 10$ and both its width and depth are larger for $b$ smaller, as implied by \eqref{eq:oscillo}. One would see more valleys at a smaller $\tau_i$.

\section{The limit $Q\rightarrow 0$}
\label{sec:Q0}

The last limit to be taken is not on the geometry, but on the family of models. The partition functions $Z_Q$ on the torus are well-defined for $Q\in (0,4]$. For models with $Q$ in this interval, the Boltzmann weight of any configuration is of the form $cQ^{\#/2}$. The power $\#$ is the number $l$ of closed loops for configurations of type $\{0\}$ and $\{a,b\}$ and $l+2$ for those of type $\mathbb Z\times \mathbb Z$. As $Q$ goes to zero, the average of the number of loops diminishes and configurations with a small number of very long loops are favored. At $Q=0$, the set of configurations is empty and, consequently, $Z_{Q=0}$ and all the partial partition functions $Z_{Q=0}(\{0\}), Z_{Q=0}(\mathbb Z\times\mathbb Z)$ and $Z_{Q=0}(\{a,b\})$ vanish. One may ask what is the homotopy of these very long loops for $Q$ very close to zero. Our intuition failed us here. This is why we explored this limit.

First note that, at $Q=0$, the expressions for the partition functions do vanish since, for $g=2$ and $e_0 = 1$, $Z(\{a,b\}) =0$  and $Z\{0\} =0$ trivially from \eqref{eq:Zab} and \eqref{eq:zed0}. This vanishing turns out to be also valid for $\tau$ away from the imaginary axis, but we shall concentrate on the case $\tau_r=0$ for the rest of this section.
The probabilities $\pi(H), H\subset \mathbb Z\times\mathbb Z$, are therefore the quotient of two quantities that tend to zero when $Q\rightarrow 0$. We first expand the partition function $Z_Q(H)$ around $Q=0$
$$Z(H)= f_0(H) + \epsilon f_1(H) + \frac{\epsilon^2} {2}f_2(H) + \cdots $$
where $\epsilon$ is a positive number such that
$$e_0 = 1-\epsilon, \qquad g = 2(1+\epsilon)\qquad \text{\rm and}\qquad Q = \pi^2 \epsilon^2.$$ As pointed earlier, $f_0$ vanishes for every subgroup $H$.

The coefficient $f_1$ for $\{a,b\}$ vanishes. Indeed, $\epsilon$ appears in \eqref{eq:Zab} only through $g$ and $e_0$ and the first derivative with respect to either at $\epsilon=0$ is easily seen to be zero. The second coefficient $f_2(\{a,b\})$ does not vanish. The second derivative may be computed by considering the variables $g$ and $e_0$ as independent first and summing their variations after. Of the three $\frac{\partial^2 Z(\{a,b\})}{\partial g^2}$,
$\frac{\partial^2 Z(\{a,b\})}{\partial g\partial e_0}$ and
$\frac{\partial^2 Z(\{a,b\})}{\partial e_0^2}$, only the third is not zero. Using again Poisson formula, we obtain
\begin{equation} \label{eq:f21}f_2(\{a,b\}) = \frac{-2\pi\tau_i}{|\eta(q)|^2(b^2\tau_i^2+a^2)^{\frac{3}{2}}}\sum_{k\in \mathbb{Z}+1/2}\left(\frac{1}{2} - \frac{2\pi\tau_ik^2}{a^2+b^2\tau_i^2}\right) q^{{k^2}/(a^2+b^2\tau_i^2)}. \end{equation}

Since $Z_Q(\mathbb Z\times\mathbb Z)=QZ_Q(\{0\})\sim \pi^2\epsilon^2Z_Q(\{0\})$ around $Q=0$, the probability $\pi(\mathbb Z\times\mathbb Z)$ can be ignored. The computation of $Z_Q(\{0\})$ is shorter as its coefficient $f_1$ is non-zero:
$$f_1(\{0\}) = \frac{-2\pi\tau_i}{|\eta(q)|^2} \sum_{m,m'\in \mathbb{Z}} \left(m^2-\frac{m'^2}{4}\right)q^{m^2+\frac{m'^2}{4}}$$
Consequently, $\pi(\{0\})$ is of order $\epsilon^0$, $\pi(\{a,b\})$ of order $\epsilon^1$ and $\pi(\mathbb{Z} \times \mathbb{Z})$ of order $\epsilon^2$. At leading order $\epsilon$, they are
$$\pi(\{0\})\sim 1,\qquad \pi(\{\mathbb{Z} \times \mathbb{Z}\})\sim \pi^2\epsilon^2$$
and
$$\pi(\{a,b\}) \sim \frac{\epsilon}{2(a^2+b^2\tau_i^2)^{\frac32} }\frac{\sum_{k\in \mathbb{Z}+1/2}(\frac{1}{2} - \frac{2\pi\tau_ik^2}{a^2+b^2\tau_i^2}) q^{{k^2}/(a^2+b^2\tau_i^2)}}{\sum_{m,m'\in \mathbb{Z}} (m^2-\frac{m'^2}{4})q^{m^2+\frac{m'^2}{4}}}.$$
Even though loops are very long in typical configurations of models with $Q$ very small, they rarely succeed in winding non-trivially around the torus. All sums in $\pi(\{a,b\})$ are related to elliptic theta functions and a compact form is
$$\pi(\{a,b\})\sim  \frac{2\epsilon}{(a^2+b^2\tau_i^2)^{\frac32}}
\frac{\frac12\theta_2(\hat q)-2\pi\tau_i \hat q
\theta_2'(\hat q)/(a^2+b^2\tau_i^2)}{4q\theta_3'(q)\theta_3(\sqrt[4]{q})-\sqrt[4]{q}\theta_3'(\sqrt[4]{q})\theta_3(q)} $$
where $\hat q=q^{1/(a^2+b^2\tau_i^2)}$, $\theta_2(q)=2\sqrt[4]{q}\sum_{0\le n<\infty}q^{n(n+1)}$ and $\theta_3(q)=\sum_{n\in\mathbb Z}q^{n^2}$.

\section{Monte Carlo simulations}
\label{sec:Numerics}

Two numerical verifications of the above results were done using Monte Carlo simulations. The first supports the claim that Pinson and Arguin's formulae hold for any $Q$'s in the interval $(0,4]$, and not only for the integers. The second measures the decay exponent $\gamma_1/b^2=\gamma_1/4$ predicted for $(1-\pi_\tau(\{1,2\}))$ in the limit $\tau=\frac12+i0^+$.

\begin{figure}[!ht]
\begin{center}
\subfigure[]{\includegraphics[width=0.45\textwidth]{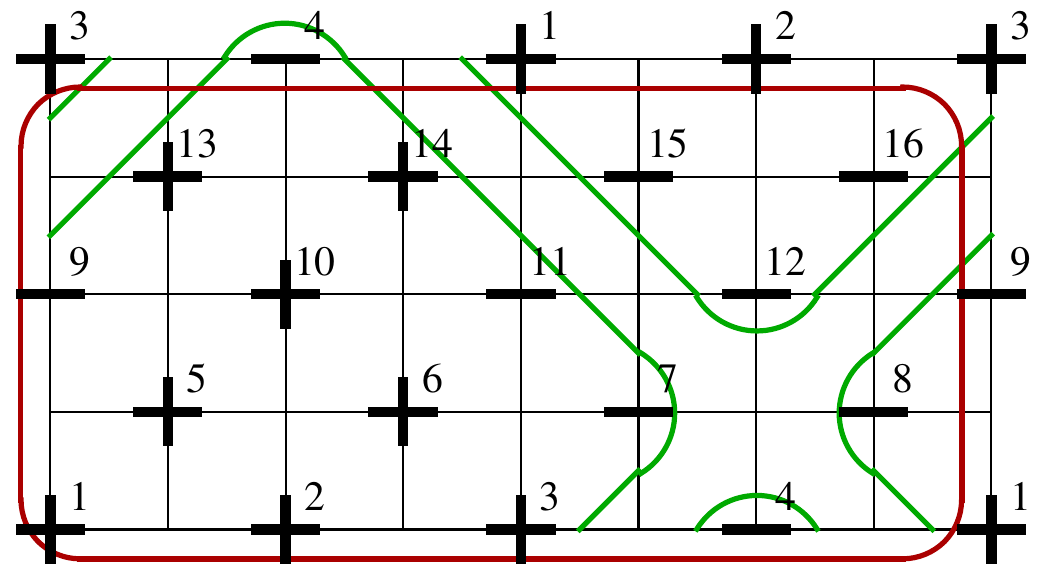}} \hfill
\subfigure[]{\includegraphics[width=0.45\textwidth]{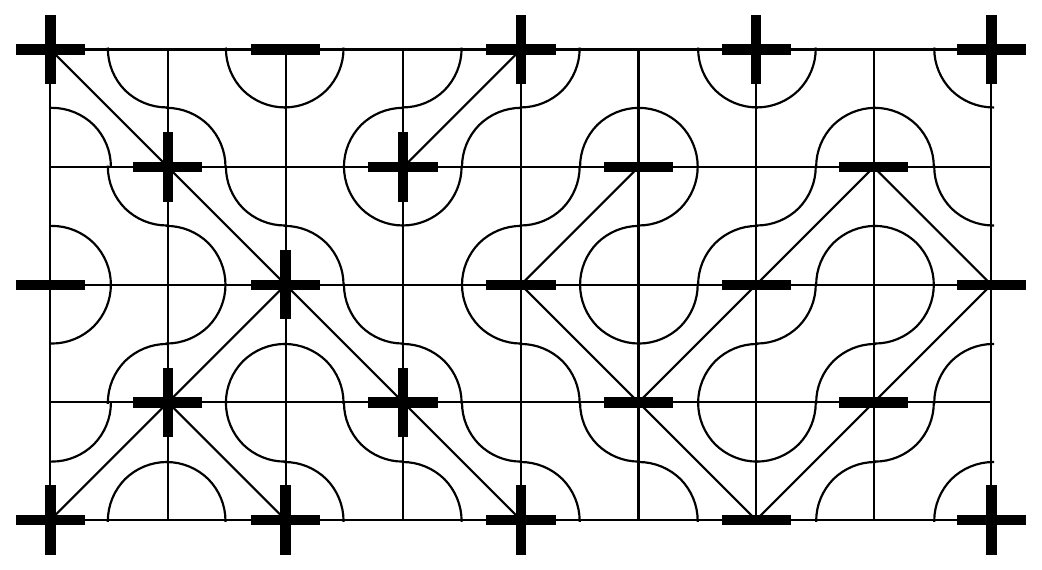}} \hfill
\caption{In (a) a $4\times 4$ spin lattice with the corresponding loop gas lattice. In (b) an admissible Fortuin-Kasteleyn configuration with the corresponding loop gas configuration.}
\label{fig:loopGas}
\end{center}
\end{figure}

Both sets of measurements were done on a family of loop gas models labeled by $Q$ that is known to describe the physics of the $Q$-Potts models when $Q$ is an integer. On figure \ref{fig:loopGas} (a), an Ising configuration on a $H\times V=4\times 4$ lattice with periodic boundary conditions is drawn obliquely. A (broken) square with rounded corners drawn at $45^o$ indicates where the $4\times 4$ lattice is. Because of the tilt, it is useful to label each spin by a number $1$ to $16$ to visualize where lie the repeated spins on the boundary. This lattice describes a torus with $\tau=i$. The basic variables of the loop gas model are the state of the smaller boxes drawn also in figure \ref{fig:loopGas} (a). The rectangle with rounded corner that lies horizontally has $h\times v=8\times 4$ such boxes. 

A Fortuin-Kasteleyn (FK) configuration, compatible with the spin configuration, has been chosen in figure \ref{fig:loopGas} (b). The FK graph is indicated by diagonals in the smaller boxes. The corresponding configuration of the loop gas is determined as follows. Note first that two of the vertices of each box is occupied by spins of the original lattice. If a FK bond is drawn between them, the state of the box is built out of two quarter-circles drawn to avoid the bond. If no FK bond is present, the two quarter-circles are drawn as to prevent a bond to appear. Note that the (loop gas) lattice of boxes $h\times v=(2H)\times H$ has sheared boundary conditions: the vertex in the bottom left is repeated in the middle of the top line. This corresponds to $\tau=\frac12+\frac i2$. The modulus for the spin lattice ($\tau=i$) and that of the loop lattice ($\tau=\frac12+\frac i2$) are distinct, but they lie in the same \slz-orbit. The Bolztmann distribution on the loop configuration is described in \cite{PearceRasmussenZuber}. In \cite{ysappjr} a simple Metropolis upgrade step is described. The number of steps sufficient for proper thermalization and the number of steps between measurements to assure statistical independence are also given there; they depend on the model, that is, on $Q$.

\subsection{Models with rational and irrational $Q$}
\label{sec:arg}

We measured the probabilities $\pi(\{0\})$, $\pi(\mathbb{Z}\times\mathbb{Z})$, $\pi(\{1,0\})$, $\pi(\{0,1\})$, $\pi(\{1,1\})$ and $\pi(\{1,-1\})$, for $Q \in \{1,\allowbreak(\frac{\sqrt{5}-1}{2})^2, \allowbreak 2,(\frac{1+\sqrt{5}}{2})^2\}$. The four values of $Q$ correspond respectively to percolation, the logarithmic minimal model $\mathcal L\mathcal M(3,5)$, the Ising model and the tricritical Ising model. The two irrational values of $Q$ test our claim that (\ref{eq:Zab}-\ref{eq:ZZ2}) apply to any $Q\in(0,4]$. The cases $Q=1$ and $2$ were also measured by Arguin \cite{Arguin} using the ``spin'' models. (The case $Q=1$ was first measured in \cite{Langlands}.) We do the measurement here using the corresponding loop gas models described summarily above. We chose to carry the simulation on a square lattice $h\times h$ with $\tau=i$. To reduce finite-size effects, the measurements were repeated for $h=4,8,16,32,64, 128$ and the estimates $\hat\pi(H)$ were obtained by making a power-law fit of the form
$$\hat\pi(H)-\hat\pi^{h\times v}(H)=C_1(h\times v)^{C_2}.$$
The results are reported in Table \ref{tab:numerics1} where the $95\%$-confidence interval is given in the form $0.1681|4$, that is $0.1681\pm 0.0004$. These are statistical errors. The agreement is excellent. Some departure from theoretical values is seen for the tricritical Ising model; this was to be expected as this model is closest to the $4$-Potts model that suffers logarithmic corrections. Results for percolation and Ising agree with~\cite{Arguin}.

\begin{table}[htbp]
\bigskip
\begin{center}\leavevmode
\begin{tabular}{|c||c|c|c|c|c|c|c|}
\hline
&&&& &&& \\[-6pt]
 Model &  $\sqrt{Q}$ & $\pi_{(\{1,0\})}$ & $\pi_{(\{0,1\})}$ & $\pi_{(\{0\})}$ & $\pi_{(\mathbb{Z}\times\mathbb{Z})}$ & $\pi_{(\{1,1\})}$ & $\pi_{(\{1,-1\})}$\\
&&&& &&& \\[-6pt]
\hline\hline
&&&& &&& \\[-6pt]

\multirow{2}{*}{${\cal LM}(3,5)$} & \multirow{2}{*}{$ \frac{1}{2} (\sqrt{5}-1)$} & $ 0.1681|4$ & $0.1682|4 $ & $0.4427|5$ & $0.1691|4 $ & $0.0258|2$ & $0.0257|2$\\
&& $0.1680$ & $0.1680$ & $0.4429$ & $0.1692$ & $0.0258$ & $0.0258$\\
&&&& &&& \\[-6pt]

\multirow{2}{*}{percolation} &  \multirow{2}{*}{$ 1$} & $0.1693|8 $ & $0.1697|8$ & $0.3094|9$ & $0.3094|9$ & $0.0211|3$ & $0.0211|3$\\
&& $0.1694$ & $0.1694$ & $0.3095$ & $0.3095$ & $0.0210$ & $0.0210$\\
&&&& &&& \\[-6pt]

\multirow{2}{*}{Ising} & \multirow{2}{*}{$ \sqrt{2}$} & $0.1466|5$ & $0.1464|5$ & $0.2259|6$ & $0.4528|7$ & $0.0141|2$ & $0.0141|2$\\
&& $0.1464$ & $0.1464$ & $0.2264$ & $0.4529$ & $0.0139$ & $0.0139$\\
&&&& &&& \\[-6pt]

\multirow{2}{*}{tric.~Ising} & \multirow{2}{*}{$ \frac{1}{2} (\sqrt{5}+1)$} & $0.1305|7$ & $0.1302|7$ & $0.1969|8$ & $0.5209|10$ & $0.0107|2$ & $0.0107|2$\\ 
&& $0.1297$ & $0.1297$ & $0.1989$ & $0.5207$ & $0.0105$ & $0.0105$\\
&&&&&&&\\
\hline

\end{tabular}
\end{center}
\caption{Numerical and theoretical probabilities for six homotopy groups, for the four models corresponding to $Q \in \{1,\allowbreak(\frac{\sqrt{5}-1}{2})^2, \allowbreak 2,(\frac{1+\sqrt{5}}{2})^2\}$.}
\label{tab:numerics1}
\end{table}

\subsection{Behavior of $\pi_\tau(\{1,2\})$ close to $\tau=\frac12$}
\label{sec:expo2}

We offer only one check of the asymptotic behavior of a $\pi_\tau(H)$ on limiting geometries. But it is a non-trivial case, $\pi_\tau(\{1,2\})$, since it probes the exponent $\gamma_1/b^2$ obtained in section \ref{sec:limit2}. This will be done for the model with $Q=(\frac12(\sqrt5-1))^2$.

The relationship between spin and loop gas lattices described earlier will play here a crucial role. For the value of $Q$ under study, there is a loop gas version, but no corresponding spin model. We keep nonetheless the name ``spin lattice'' for the lattice drawn obliquely in figure \ref{fig:loopGas} (a) and use capital letters to give its size. (Note that the letter $H$ is also used for the subgroup of the holonomy group. Hopefully this will not cause any confusion.) We aim at measuring various probabilities $\pi_\tau(H)$ for spin lattices of size $H\times V$ with $V=20$ and $H=V/\tau_i$ for small $\tau_i$, that is, for $H\gg V$. We choose $H=20, 40, 60, 80, 120, 160, 320, 640, 960$ and $1280$. The corresponding value of $\tau=\tau_r+i\tau_i$ are $\tau_r=\frac12$ and $1/\tau_i=1, 2, 3, 4, 6, 8, 16, 32, 48$ and $64$. To account for $\tau_r=\frac12$, the bottow row of the spin lattice has to be shifted to the right by $C=H/2$ sites before being glued to the top row. 

For these sizes of spin lattices, the corresponding loop lattices have size $h\times v$ (with $h$ and $v$ in small letters) given by
$$h=\frac{2V \text{\rm\ lcm} (C+V, H)}{C+V},\qquad v=\frac{2HV}{h}$$
where $\text{\rm lcm}$ denotes the least common multiple. A shift $c$, similar to $C$ for the spin lattice, is necessary for the loop gas lattice. This shift $c$ is obtained by solving
$$\alpha(-V+C)+\beta H=c,\qquad \alpha(V+C)+\beta H=v$$
for $\alpha, \beta$ and $c$ under the constraints $\alpha, \beta, c\in\mathbb Z$ and $c\in[0,h)$. As an example, the loop lattice $h=1280, v=20, c=620$ corresponds to the spin lattice $H=640, V=20, C=320$. The samples are of $2\times 10^6$
configurations
for the six smallest lattices and of $10^6$ for the four largest.
\begin{figure}
 \centering
  \includegraphics[width=0.80\textwidth]{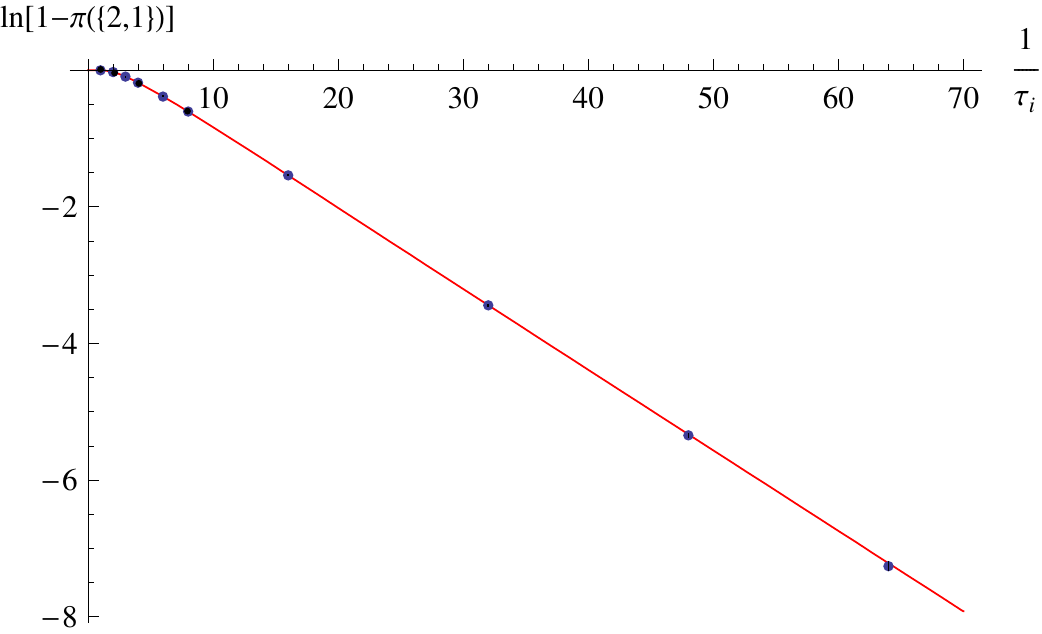}
 \caption{Numerical values for $\pi_{(\{1,2\})}$ are plotted, with the theoretical curve and the linear fit obtained using the five largest lattices. The size of the dots are larger than the statistical errors.}
 \label{fig:numericlimit2}
\end{figure}

The results appear in Table \ref{tab:numerics2}. In all cases the agreement for $\pi(\{1,2\})$ is excellent, up to three or four digits. This is remarkable considering that one of the lattice linear sizes is very small. Indeed the six largest (loop) lattices have $v=20$ and finite-size effects should be present. It is also welcome since the measurement of $\gamma_1$ requires to take the logarithm of $(1-\hat\pi(\{1,2\}))$. The slope on figure~\ref{fig:numericlimit2} should be $\frac{2\pi \gamma_1}{b^2}$ at large $1/\tau_i$. In the present case, $b=2$ and $\gamma_1=\frac3{40}= 0.075$. Using only the six largest lattices, we extract $\hat\gamma_1=0.0756 \pm 0.0005$, a reasonable agreement,
as again the error does not include finite-size effects.

\begin{table}[h]
\bigskip
\begin{center}\leavevmode
\begin{tabular}{|c||c|c|c|c|c|c|c|c|}
\hline
& &&&&&&& \\[-6pt]

$1/\tau_i$ & $\pi_{(\{0\})}$ & $\pi_{(\mathbb{Z}\times\mathbb{Z})}$ & $\pi_{(\{1,0\})}$ & $\pi_{(\{0,1\})}$ & $\pi_{(\{1,1\})}$ & $\pi_{(\{1,-1\})}$ & $\pi_{(\{2,1\})}$ & $\pi_{(\{1,2\})}$\\ 

&&&&&&&& \\[-6pt]
\hline\hline
&&&&&&&& \\[-6pt]

\multirow{2}{*}{$1$} & $0.4477|7$ & $0.1710|6$ & $0.1663|6$ & $0.1051|5 $ & $0.1045|5$ & $0.00244|7$ & $0.00235|7$ & $0.00065|4$\\
                     & $0.4480 $  & $0.1711 $  & $0.1661 $  & $0.1047 $   & $0.1047 $   & $0.00242 $  & $0.00242 $  & $0.00059 $ \\

&&&&&&&& \\[-6pt]

\multirow{2}{*}{$2$} & $0.4434|7$ & $0.1690|5$ & $0.0259|3$ & $0.1679|6$ & $0.1676|6$ & $0.00010|14$ & $0.000095|14$ & $0.0257|3$\\
                     & $0.4429$  & $0.1692$  & $0.0258$  & $0.1680$     &  $0.1680$  &  $0.00009$    & $0.000090$   & $ 0.0258 $ \\

&&&&&&&& \\[-6pt]

\multirow{2}{*}{$3$} & $0.4484|10$ & $0.1710|8$ & $0.0039|2$ & $0.1430|7$ & $0.1427|7$ & $0.000001|2$ & $0.000001|2$ & $0.0892|6$\\
                     & $0.4486$    & $0.1714$   &  $0.0039$  & $0.1424$  &  $0.1424  $ & $0.000002$   & $0.000002$   & $0.0894$ \\

&&&&&&&& \\[-6pt]

\multirow{2}{*}{$4$} & $0.4485|7$ & $0.1709|6$ & $0.00060|4$ & $0.1047|5$ & $0.1045|5$ & $NE$ & $NE$ & $0.1660|6$\\
                     & $0.4480$   & $0.1711$   & $0.00059$   & $0.1047$  &  $0.1047$  &  $0$   & $0$   & $0.1661 $ \\

&&&&&&&& \\[-6pt]

\multirow{2}{*}{$6$}& $0.4107|10$ & $0.1568|8$ & $0.000013|8$ & $0.0529|5$ & $0.0528|5$ & $NE$ & $NE$ & $0.3183|10$\\
                    & $0.4105$    & $0.1568$  &  $0.000015$   & $0.0527$  & $ 0.0527$  &  $0$   & $0$   & $0.3186 $ \\

&&&&&&&& \\[-6pt]

\multirow{2}{*}{$8$} & $0.3509|7$ & $0.1339|5$ & $NE$ & $0.0270|3$  & $0.0267|3$ & $NE$ & $NE$ & $0.4533|8$\\
                     & $0.3511$  &  $0.1341$  &  $0$   &$0.0268 $  &  $0.0268 $ &  $0$   & $0$   & $0.4529 $ \\

&&&&&&&& \\[-6pt]

\multirow{2}{*}{$16$} & $0.1510|7$ & $0.0577|5$ & $NE$ & $0.00219|9$ & $0.00213|9$ & $NE$ & $NE$ & $0.7850|8$\\
                      & $0.1508$   & $0.0576$   & $0$   & $0.00220$   & $0.00220$   & $0$   &  $0$  & $0.7850 $ \\

&&&&&&&& \\[-6pt]

\multirow{2}{*}{$32$} & $0.0231|3$ & $0.0089|2$ & $NE$ & $0.000018|9$ & $0.000017|8$ & $NE$ & $NE$ & $0.9679|4$\\
                      & $0.0232$  &  $0.0088$  &  $0$   & $0.000023$   & $0.000023$   & $0$   & $0$   & $0.9679 $ \\

&&&&&&&& \\[-6pt]

\multirow{2}{*}{$64$} & $0.00050|5$ & $0.00020|3$ & $NE$ & $NE$ & $NE$ & $NE$ & $NE$ & $0.99930|5$\\
                      & $0.00053$  & $0.00020$   & $0$   & $0$   & $0$   & $0$   & $0$   & $0.99927 $ \\

&&&&&&&&\\
\hline

\end{tabular}
\end{center}
\caption{Numerical and theoretical probabilities for ${\cal LM}(3,5)$. The probabilities $\pi_{(\{2,-1\})}$ and $\pi_{(\{1,-2\})}$ are not included because their numerical and theoretical values are $\lesssim 10^6$. Theoretical values are set to 0 when $\lesssim 10^6$ and $NE $ indicates that no such event was recorded.} 
\label{tab:numerics2}
\end{table}

\section*{Appendix}

We prove here the special values (\ref{eq:c0}-\ref{eq:c1}) of the function $C(d,e_0)$. Its definition is
\begin{align*}C(d,e_0)&=\sum_{d_2|d} \cos(\pi e_0d_2)\mu(d/d_2)\\
\intertext{but, if the sum is done over $d_3=d/d_2$, it can also be written as}
&=\sum_{d_3|d}\cos(\pi e_0d/d_3)\mu(d_3).
\end{align*}
Because the cosine function is even and periodic, the function $C(d,e_0)$ satisfies
$$C(d,e_0)=C(d,-e_0)=C(d,e_0+2).$$
The key relation for proving (\ref{eq:c0}-\ref{eq:c1}) is
\begin{equation}\label{eq:key} C(d,ke_0)=C(kd,e_0)+\delta_{k\wedge d,1}C(d,e_0),\qquad \text{\rm for $k$ prime}.\end{equation}
We first prove it.

By definition 
\begin{align*}
C(d,e_0) &= \sum_{d_2|d} \cos\left( \frac{\pi e_0}k \cdot\frac{kd}{d_2} \right)\mu(d_2)\\
&=\sum_{d_2|kd}\cos\left( \frac{\pi e_0}k\cdot \frac{kd}{d_2}\right)\mu(d_2)-
{\sum}'\cos\left( \frac{\pi e_0}k\cdot \frac{kd}{d_2}\right)\mu(d_2).\end{align*}
Bu summing over all divisors of $kd$ instead of those of $d$ only, the first sum of the second line has added terms; the sum ${\sum}'$ is over these spurious terms and restores therefore the equality with the previous line. Suppose $k$ is prime. If $d$ has $k$ among its prime factors, all the divisors $d_2$ of $kd$ that are not divisors of $d$ contain $k^2$ as factors. Their Moebius factor $\mu(d_2)$ is then $0$ and the sum ${\sum}'$ vanishes. If $k$ is not a prime factor of $d$, then all $d_2$ in the sum ${\sum}'$ are of the form $d_2=kd_3$ with $d_3$ a divisor of $d$. Then
\begin{align*}
{\sum}'\cos\left( \frac{\pi e_0}k\cdot \frac{kd}{d_2}\right)\mu(d_2)&=
\sum_{d_3|d} \cos\left( \frac{\pi e_0}k\cdot \frac{kd}{kd_3}\right)\mu(kd_3)\\
&=- C(d,e_0/k).
\end{align*}
We have thus proved
\begin{equation}\label{eq:keyk}C(d,e_0)=C(kd,e_0/k)+\delta_{d\wedge k,1} C(d,e_0/k).\end{equation}
Equation \eqref{eq:key} follows if $e_0$ is replaced by $ke_0$. Both forms are useful.

The first identity in (\ref{eq:c0}-\ref{eq:c1}) is almost trivial. But it shows how to use \eqref{eq:keyk}. Suppose $d$ has a repeated prime factor, say $d=k^2d'$. Then $k\wedge (kd')=k$ and the identity \eqref{eq:keyk} gives $C(kd',0)=C(k^2 d',0)+\delta_{k\wedge (kd'),1}C(kd',0)=C(d,0)$. The $d$'s to be studied are therefore those with only distinct prime factors. Suppose that $d\neq 1$ has $l$ such factors. In the definition of $\mu(d_2)$, only the number of prime factors is important and, if the sum over divisors is replaced by a sum over the number of prime factors in these divisors, $C(d,0)$ becomes
$$C(d,0)=\sum_{d_2|d}1\cdot \mu(d_2)=\sum_{i=0}^l (-1)^i{l\choose i}=(1-1)^l=0.$$
Finally $C(1,0)=\mu(1)=1$, which proves \eqref{eq:c0}.

The last identity \eqref{eq:c1} is the next to be proven. The periodicity of $C$ simplifies its study. If $k$ is an odd prime, then $C(d,k)=C(d,1)$. For a given $d$, choose an odd prime $k$ such that $k\wedge d=1$. Then \eqref{eq:key} gives
$$C(d,1)=C(d,k)=C(dk,1)+C(d,1),$$
proving $C(kd,1)=0$. This states that
\begin{equation}C(d',1)=0\label{eq:cdp1}\end{equation}
if $d'$ has a non-repeated odd prime among its prime factors. 

Like above, suppose that $d$ has a repeated odd prime factor, $d=k^2d'$. Then \eqref{eq:key} and periodicity give
$$C(d,1)=C(kd',k)=C(kd',1).$$
Removing further factors $k$ if necessary, one can bring these cases back to \eqref{eq:cdp1}. The only remaining cases are $d$ a power of $2$ and $d=1$. If $d=2^n, n\ge 1$, then \eqref{eq:key} and periodicity give
$$C(2d,1)=C(d,2)=C(d,0)=\delta_{d,1}=0.$$
A direct calculation gives $C(2,1)=2$ and $C(1,1)=-1$, proving \eqref{eq:c1}.

Let $e_0=\frac12$ now. If $d$ is odd, then all its divisors $d_2$ will also be and then $\cos(\pi e_0 d_2)=0$ and $C(d,\frac12)=0$. The periodicity and evenness of $C$ implies also that $C(d,\frac{k}2)=C(d,\frac12)$ for $k$ odd. This allows to use again \eqref{eq:key} efficiently. For $k$ odd
$$C(d,{\textstyle{\frac12}})=C(d,{\textstyle{\frac{k}2}})=C(kd,{\textstyle{\frac12}})+
\delta_{k\wedge d,1}C(d,{\textstyle{\frac12}}).$$
From this point on, the argument is similar to that for $C(d,1)$. The proof of the last two cases ($e_0=\frac13$ and $\frac23$) uses no new argument and will be omitted.

These special cases might lead one to think that, for any rational $e_0$, the set 
$\{d\in\mathbb N^*\,|\, C(d,e_0)\neq0\}$
is finite. This is false. The cases $e_0=0,\frac13, \frac12,\frac23,1$ are exceptional in this sense. It is intriguing to note that these values of $e_0$ are precisely those corresponding to the Potts models with $Q=4,3,2,1,0$ respectively. (The limiting value $Q=0$ corresponds to dense polymers.)

%
%

\section*{Acknowledgements} 
We thank John Cardy, Robert Ziff and Andrew Granville for helpful discussions. AMD holds a scholarship and YSA a grant of the Canadian Natural Sciences and Engineering Research Council.
AMD also holds a scholarship of the Fonds Quebecois de la Recherche sur la Nature et les Technologies. This support is gratefully acknowledged.

%
%


\end{document}